\documentclass{aa}
\usepackage{graphicx}
\usepackage{mathptmx}
\usepackage[T1]{fontenc}
\usepackage{ae,aecompl}

\usepackage{amsmath}    % Advanced maths commands
\usepackage{amssymb}    % Extra maths symbols

\usepackage{lineno}

\usepackage{hyperref}
\begin{document} 

%\linenumbers

   \title{Forbidden planetesimals}

   \author{L. Sch\"onau
          \and
          J. Teiser
          \and
          T. Demirci
           \and
          K. Joeris
           \and
          T. Bila
           \and
          F.C. Onyeagusi
           \and
          M. Fritscher
           \and
          G. Wurm
          }

   \institute{University of Duisburg-Essen, Faculty of Physics, Lotharstr. 1, 47057 Duisburg, Germany\\
              \email{laurent.schoenau@uni-due.de}      
             }

   \date{Received; accepted }

% \abstract{}{}{}{}{} 
% 5 {} token are mandatory

\abstract {Planetesimals are born fragile and are subject to destruction by wind erosion as they move through the gas of a protoplanetary disk. In microgravity experiments, we determined the shear stress necessary for erosion of a surface consisting of 1 mm dust pebbles down to 1 Pa ambient pressure. This is directly applicable to protoplanetary disks. Even pebble pile planetesimals with low eccentricities of 0.1 cannot survive inside of 1 au in a minimum-mass solar nebula, and safe zones for planetesimals with higher eccentricities are located even farther out.

}


   \maketitle
%
%-------------------------------------------------------------------

\section{Introduction}
\label{sec:introduction}

Planet formation spans a huge variety in physical mechanisms, timescales, and spatial scales. Models are therefore subdivided into quite a number of steps \citep{Blum2008, Wurm2021, Johansen2014}. Working their way from dust to planets, these different steps are often related to certain size ranges with their specific evolution paradigms.
The first of these paradigms is that initially micron-sized dust in protoplanetary disks collides, sticks together, grows, and becomes compressed, until eventually, millimeter-sized, compact dust aggregates are formed \citep{Blum2000, Wurm1998, Blum2000a, Kruss2017}. The exact size is debatable, but with growth and compaction at the millimeter-scale, collisions eventually enter a regime of more or less elastic rebound. This is known as the bouncing barrier \citep{Zsom2010, Windmark2012, Kelling2014, Kruss2016, Demirci2017}. Further simple growth via ongoing sticking collisions where van der Waals forces provide the binding is challenging from here, hence the name bouncing barrier.

In view of rebounding collisions, collisional charging (tribocharging) might become important. It has been shown in recent years that collisional charging can induce a new growth phase, now moderated by long-range Coulomb forces \citep{Steinpilz2019, Jungmann2018, Jungmann2021, Lee2015}. In this way, clusters of aggregates of decimeter-size might grow \citep{Teiser2021}. The individual entities in these clusters are still the bouncing-barrier size dust aggregates that are not restructured into a larger aggregate. However, the larger clusters now have overall aerodynamic properties of a larger body \citep{Schneider2021}. These ideas about collisional charging in the context of planet formation are not a paradigm yet and are still evolving, but they might be a welcome if not necessary bridge in size scale to the next phase. Particles of decimeter-scale rather than millimeter are concentrated in varying ways, from pressure maxima to drag instabilities and especially streaming instabilities \citep{Dittrich2013, Carrera2022, Chiang2010, Johansen2006}. If this concentration works well, that is, to a level of high densities, the solids might eventually collapse gravitationally to produce a larger body or planetesimal \citep{Johansen2014}. These objects only collapse gently, however, and it has been shown that the gravitational compression in this phase for small planetesimals is not strong enough to restructure them. Therefore, these objects become pebble piles with little self-gravity, and pebbles are dust aggregates \citep{Blum2017, Wahlberg2017}.

Due to this history, with little gravity and sticking, pebble pile planetesimals are very fragile. Any distortion might disassemble them again. This is especially true for gas drag, which is omnipresent in protoplanetary disks. As planetesimals are embedded within, the least they feel is a head wind \citep{Weidenschilling1977}. With about 50 m/s, this is rather high in favor of erosion. On the other hand, the pressure at submillibar is rather low, which reduces the shear stress imposed by gas drag significantly. On the destructive side, the almost negligible gravity might again make it easier to disassemble the body. These are clearly rather extreme conditions, and little experimental data or evidence is available on the impact of this delicate balance between opposing forces. 

The closest related research in which experimental data have been gathered was made on erosion on Mars, where gravity is slightly reduced to 1/3 g and the atmospheric pressure is a few millibar. 
Wind tunnel experiments have been conducted in the Mars-relevant regimes of low pressure and/or low gravity \citep{Greeley1980,White1987, Musiolik2018, Kruss2019, Demirci2019}.
Still, these data are orders of magnitude different from protoplanetary disk values.
\citet{Paraskov2006} studied dust erosion at low pressures with planet formation in mind, but their micron-sized particles were no pebble piles. \citet{Demirci2020} were the first to study erosion of millimeter-size glass beads simulating pebbles under low gravity of 0.01 g and at a pressure as low as 10 Pa. At these pressures, the flow changes from continuum to free molecular with respect to the individual grains as the mean free path of the molecules $\lambda$ exceeds the particle diameter $d$. While other dependences on gravity and surface energy might still hold and have been deduced before (e.g., \citet{Shao2000}),
the low-pressure range requires a correction factor $f_c$, as  shown by \citet{Demirci2020}. In any case, the important parameter for wind erosion is the threshold shear stress.
%--------------------------------------------------------------------
\section{Shear stress}
\label{sec: Shear stress}

So far, threshold shear stress with respect to the onset of erosion on a planetesimal or small body surfaces can be characterized as \citep{Demirci2020} 
\begin{eqnarray}
\label{eq:tau_erosion_final}
    \tau_\mathrm{t}=\alpha f_\mathrm{c} \left(\frac{\rho_\mathrm{p} g d}{9} +\frac{\gamma_\mathrm{e}}{d}  \right)\\
\nonumber    f_\mathrm{c}=1+ \frac{\mathrm{2 \cdot Kn}}{\beta} \left(1.257 + 0.4 e^{- \frac{0.55 \beta}{\mathrm{Kn}}}  \right)\\
\nonumber    \mathrm{Kn}=\frac{\lambda}{d}=\frac{k_\mathrm{b}}{\sqrt{2} \pi d_\mathrm{m}^2 R_\mathrm{g} \rho_\mathrm{g} d}.
\end{eqnarray}
Here, particle or surface properties are the particle density $\rho_\mathrm{p}$, particle diameter $d$, the effective surface energy $\gamma_\mathrm{e}$, and $\alpha$, which is a friction coefficient. The gravitational acceleration is $g$. The factor $f_\mathrm{c}$ holds the influence of the gas regime, namely the Knudsen number Kn and $\beta,$ which might be interpreted as a correction factor as particles are not within a free flow, but on a surface. The Knudsen number depends on the gas density $\rho_\mathrm{g}$.
Furthermore, $k_\mathrm{b} = 1.38 \cdot 10^{-23} \, \rm J \, K^{-1}$ is the Boltzmann constant, $d_\mathrm{m} = 0.37 \cdot 10^{-9} \, \rm m$ is the molecule diameter, and $R_\mathrm{g} = 287.058\, \rm J \, kg^{-1} \, K^{-1}$ is the specific gas constant for dry air. The gas density can be determined from the pressure $ p $ using the ideal gas equation $ p = \rho_\mathrm{g} R_\mathrm{g} T $ with the temperature $T = 290 \,\mathrm{K}$.

Free parameters that might change with the sample are $\alpha$, $\beta$, and $\gamma_\mathrm{e}$. The best values based on experiments with glass beads so far were
$\alpha = 6.4 \cdot 10^{-3}$, $\gamma_\mathrm{e}= 7.3 \cdot 10^{-5} \,\mathrm{J}\,\mathrm{m}^{-2}$, and
$\beta = 0.67$ (for large glass beads) \citep{Demirci2020}. 

\subsection*{Erosion in protoplanetary disks}

When this is applied to protoplanetary disks, there are a few general consequences. First,
pebble-pile planetesimals are destroyed on certain orbits. This has recently been studied by a number of authors \citep{Rozner2020, Schaffer2020, Cedenblad2021, Demirci2020b, Demirci2020}. Due to the head wind in a typical disk, where ambient pressure decreases with distance, there should be a planetesimal-free region within the inner 1 au as there is always a head wind. If planetesimals consist of compressed dust and are no pebble piles, this region is small \citep{Paraskov2006}. For pebble piles, however, it can be quite significant and extend to several tenths of  1 au. This is therefore really important for planet formation. 
%What would happen to the pebbles then that do not make it into a planetesimal for good, is a different question, but if later phases of planet formation rely on a reservoir of planetesimals this might be rather problematic in the inner 1 au.

Another consequence is that planetesimals might preferentially orbit on circular orbits. 
Before larger planetary bodies are present, for the first generation of planetesimals, this might not be much of an issue as planetesimals would not be on eccentric orbits. However, after the first larger oligarchs formed, planetesimals can be scattered into eccentric orbits \citep{Ida1990, Higuchi2006}. Larger bodies have to be present before the gaseous disk dissolves in order to evolve into gas giants, and disks with embedded planets have been observed by now \citep{Keppler2018}. Gas drag will again decrease the eccentricities, but even in a gaseous disk, it will take several orbits for a planetesimal before the orbit might be circularized  \citep{Demirci2020b}.
As long as it moves on an eccentric orbit, a planetesimal will experience much higher wind speeds and gas densities at periastron. These will disassemble it at least partially on each orbit \citep{Cedenblad2021, Schaffer2020, Paraskov2006, Demirci2020b}. 

There is another interesting implication here as this might transport material in the form of large planetesimals from  outside of 1 au and redistribute it to pebbles inside of 1 au. So far, these ideas have not been exploited in detail. However, with a now much denser region, planetesimals on circular orbits could form, which might lead to a boost of planetesimal formation within the inner 1 au at the expense of eccentric planetesimals.

Finally, a further consequence is the erosion of planetesimals with respect to pebble accretion somewhat later in the evolution toward planets
\citep{Demirci2020c}. Even if planetesimals in the outer region are stable during their formation, when they later interact with a forming giant planet or planet with an atmosphere, the increased gas density will again allow a disassembly of the planetesimals.
This has a large impact on the mass balance of accretion \citep{Demirci2020c}.

With these applications in mind, we continued with experiments in microgravity to
approach the conditions of protoplanetary disks ever more realistically.
Here, we specifically use dust aggregates as pebbles instead of glass beads. We also improved the setup so that we were able to quantify the erosion threshold at lower pressure, that is, as low as 1 Pa.

\section{Parabolic flight experiments}
\label{sec:microgravityexperiments}

\subsection{Particle sample}

One important motivation for this current work was to no longer simulate the planetesimal surface with glass beads. Instead, we moved to real dust aggregates that were produced quite similarly to what we would expect in protoplanetary disks, namely growing the dust aggregates until they reach the bouncing barrier. To do this, we started with $\rm SiO_2$ (quartz) dust here. Importantly, the individual dust grains were a few micrometers in size and therefore in the range expected in protoplanetary disks. In detail, our sample followed a log-normal distribution centered at 3 micrometers (see, e.g., \citet{Kruss2018} for the size distribution).

\begin{figure}
    \centering
    \includegraphics[width=\columnwidth]{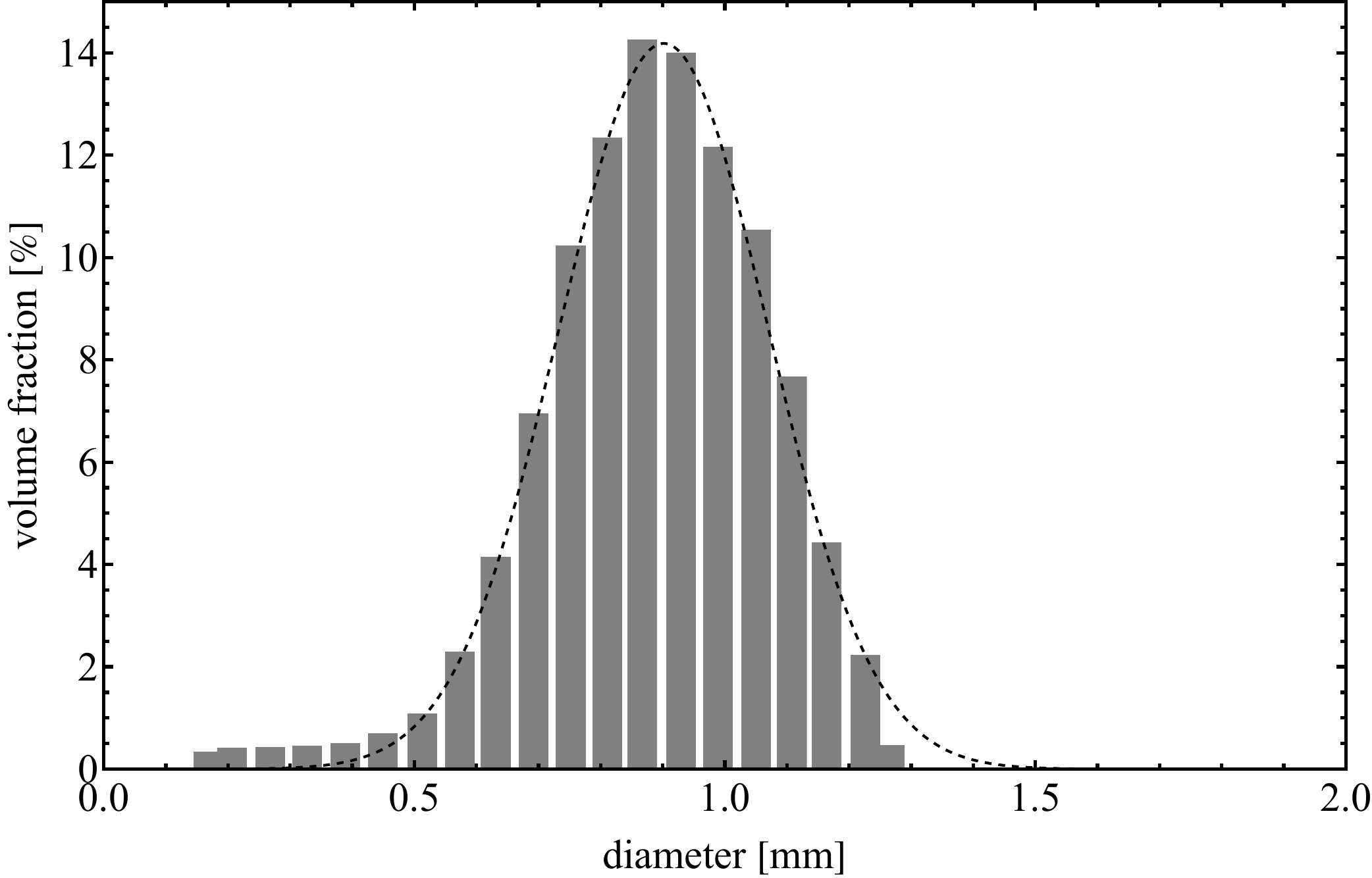}
    \caption{\label{fig:size} Size distribution of dust aggregates.}
\end{figure}

\begin{figure}
    \centering
    \includegraphics[width=\columnwidth]{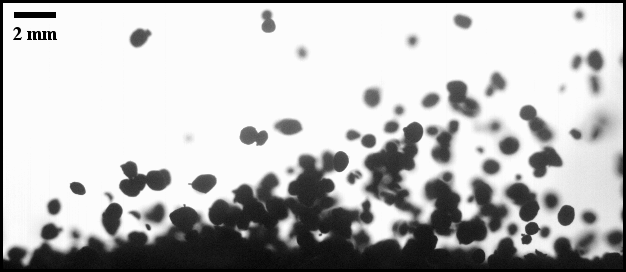}
    \caption{\label{fig:sotheylook} Image of the dust aggregates from an in-flight video.}
\end{figure}

To produce the initial dust aggregates for the erosion experiments, the dust was placed on a vibrating plate. Dust grains now
stuck together in collisions, grew, and were compacted until they reached about 1 mm in size. They now no longer grew further, but bounced off each other. After this process, the dust aggregates have the size distribution shown in Fig. \ref{fig:size}.
The volume filling factor of the dust aggregates is about 0.33, as determined by measuring the size and mass of numerous individual aggregates. An image of the eroding dust aggregates from an in-flight video is shown in Fig. \ref{fig:sotheylook}.
Overall, the aggregates are a factor 2 larger than the largest glass beads used before in \citet{Demirci2020}. They have different sticking properties because the surface is now composed of dust grains rather than being a flat glass surface, and the size distribution is slightly broader. These differences  will likely influence the free parameters of Eq. \ref{eq:tau_erosion_final}. 

\subsection{Experimental setup}

\begin{figure}
    \centering
    \includegraphics[width=\columnwidth]{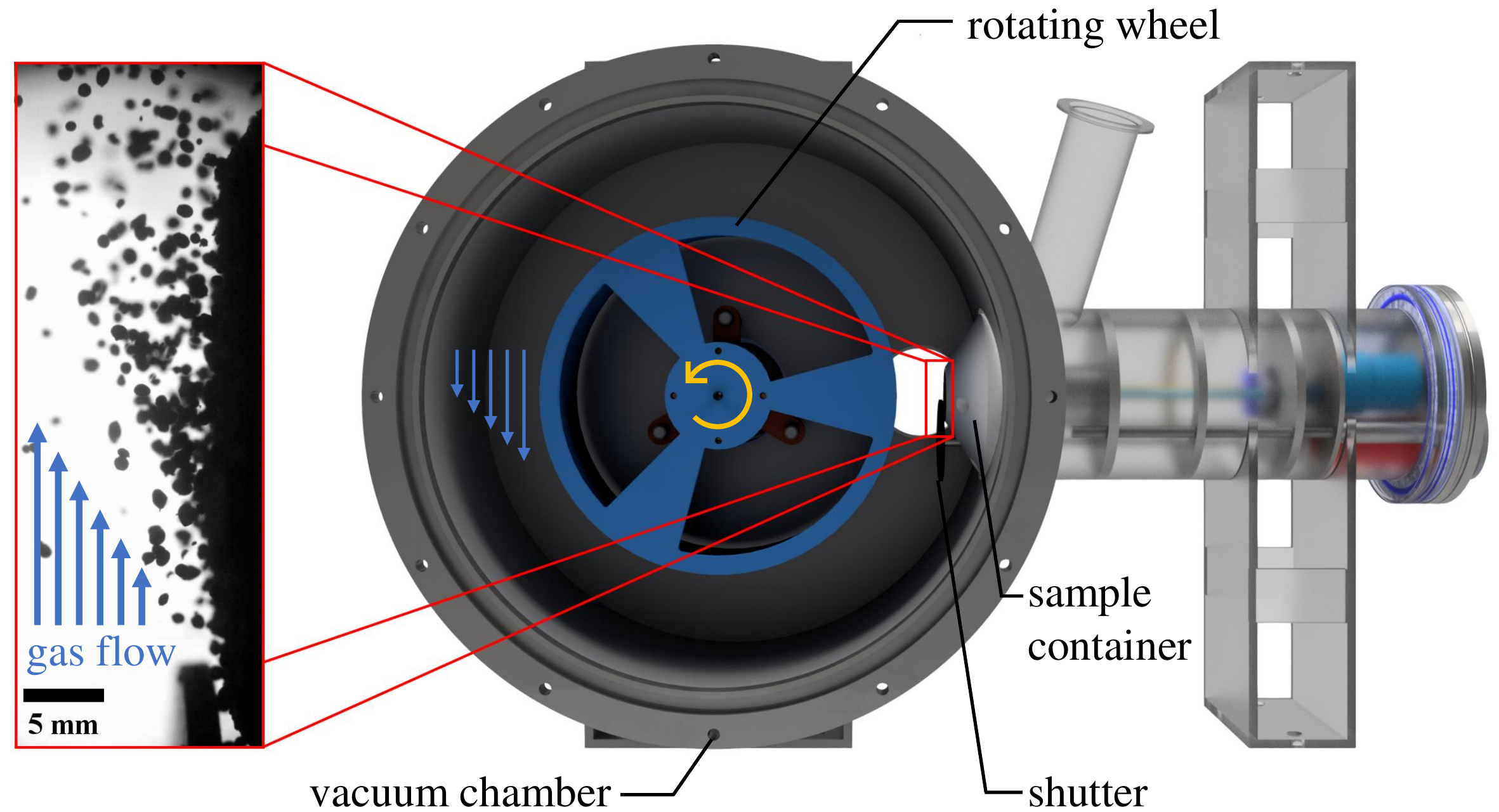}
    \caption{\label{fig:setup} Sketch of the experimental setup adapted from \citet{Demirci2020}. Erosion of a particle bed (right side inside the cylindrical vacuum chamber) is generated by a shear flow.}
\end{figure}

Figure \ref{fig:setup} shows a schematic of the experimental setup, which is the same as was used by \citet{Demirci2020}. In a cylindrical vacuum chamber, a second smaller cylinder rotated at adjustable frequencies up to 200 Hz to create a shear flow. This design allowed the generation of a laminar wind profile over a sample bed at low pressures down to 1 Pa with a maximum transversal cylinder-wall speed of 125 m/s. The dust sample was placed in a sample container attached to the side of the vacuum chamber. A lifter mechanism was used to move the surface of the dust bed into the wind. In addition, a shutter ensured that the sample bed was only exposed to the wind when the gravitational conditions were within a predefined range.

In order to simulate these gravitational conditions with this experiment, it was developed to be used in an aircraft that performed parabolic flights. We participated in the 38th DLR parabolic flight campaign in February 2022 on board Novespace's \textit{Airbus A310 Zero-G}. During the microgravity phase of a parabola, the residual gravity (g-jitter) has a characteristic profile. The g-jitter in flight direction is greater than the g-jitter in the other two directions only for a short period of 3 to 4 seconds, so that the total gravity is mainly directed into the sample container. As long as this is the case, the shutter is open and the sample bed is exposed to the wind. During the parabola, images of the dust sample are recorded by a side-mounted camera at a frame rate of 2002 fps. The pressure in the chamber can be varied as needed. For a more detailed description of the experiment itself and the procedure during a parabola, we refer to \citet{Demirci2020}.

\subsection{Shear flow}

The earlier experiments by \citet{Demirci2020} showed that at submillibar pressure, the gas flow within the experiment is laminar. The flow velocity increases linearly with height until it reaches the speed of the central rotating wheel at its surface. We verified these characteristics again by tracking dust grains and deducing local gas speeds from them in analogy to \citet{Demirci2020}. 

The details are not shown here as they do not provide new results. However, we again find a linear dependence of gas speed $u$ on height $h,$ in agreement with a simple laminar shear flow. From this, we can deduce the shear stress at a given speed of the wheel $v_\mathrm{w}$ as  
\begin{equation}
\label{eq:shearing2}
    \tau = \eta \frac{v_\mathrm{w}}{h_\mathrm{w}}.
\end{equation}
Here, $\eta = 18 \,\mathrm{\mu}\,\mathrm{Pa}\,\mathrm{s}$ is the viscosity of air, and $h_\mathrm{w} = 0.033 \, \rm m$ is the height of the wheel over the dust surface. As dust surface, we defined the position at which a linear fit to the measured velocities would be zero, which is in agreement with the observed surface. By successively increasing $v_\mathrm{w}$ during the experiment until aggregates were clearly displaced by the wind, we were able to determine the threshold shear stress $ \tau_\mathrm{t} $ for erosion at different ambient pressures.

\section{Results}

\begin{figure}
        \centering
        \includegraphics[width=\columnwidth]{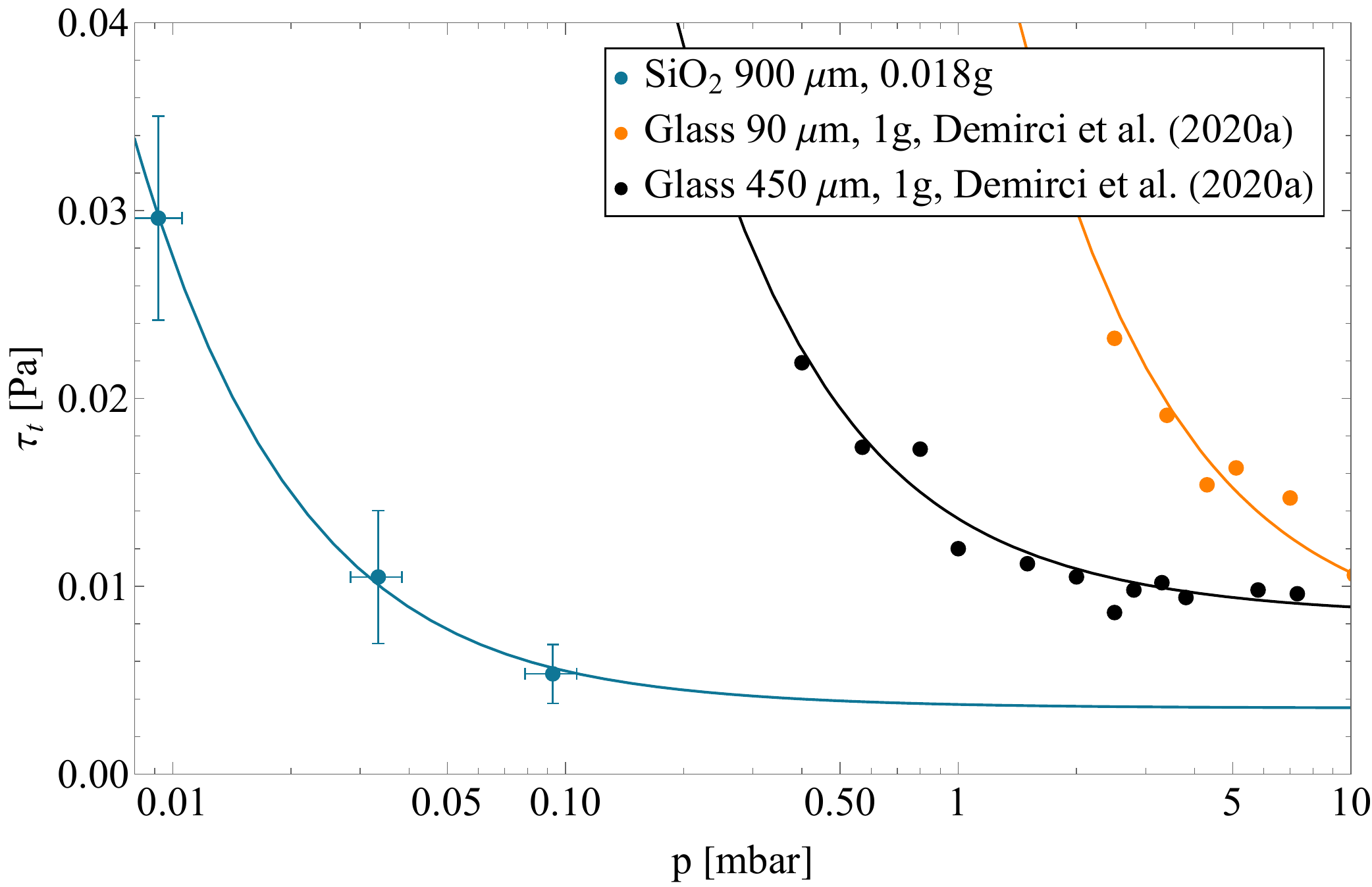}
        \caption{\label{results} Threshold shear stress $ \tau_\mathrm{
        t} $ over ambient pressure for the dust aggregates at 0.018 g and for glass spheres at 1 g from earlier experiments. The solid lines are fits of equation \ref{eq:tau_erosion_final} to the data. The free parameters for dust aggregates, received from the fit, can be found in equation \ref{eq:werte}, and the free parameters for the larger glass spheres can be found at the end of section \ref{sec: Shear stress}. For more details on the glass sphere measurements, we refer to \cite{Demirci2020}. } 
\end{figure}

One main goal was to determine erosion thresholds for shear stress at pressures as low as possible for dust aggregates to place the parameters in the range of protoplanetary disks. The onset of erosion was visible at a pressure of 1 Pa, an order of
magnitude lower than before and well within the transition regime
between free molecular flow and continuum flow.
These threshold values are shown in Fig. \ref{results}.

The expected pressure dependence according to Eq. \ref{eq:tau_erosion_final} was fit to the data, or
\begin{equation}
    \tau_\mathrm{t} = f_\mathrm{c}  \cdot \tau_\infty  
    \label{nixneues}
,\end{equation}
where $\tau_\infty$ is the shear stress for very high pressures, where $f_\mathrm{c} = 1,$ and $f_\mathrm{c}$ with its $Kn / \beta$ dependence characterizes the increase at low pressure. 
According to Eq. \ref{eq:tau_erosion_final}, it is
\begin{equation}
\label{shortcut}
    \tau_\infty =     \alpha \left(\frac{\rho_\mathrm{p} g d}{9} +\frac{\gamma_\mathrm{e}}{d}  \right).
\end{equation}
While for the microgravity experiments, gravity is likely lower than adhesion, they do not allow specifying $\alpha$ or $\gamma_\mathrm{e}$ independently. We therefore also carried out experiments for the threshold shear stress for dust aggregates under Earth gravity at 1 g. In this case, the gravity term dominates, that is, the adhesion term can be neglected and $ \alpha$ can be calculated if $\tau_\infty $ is known. For the 1 g experiment with dust aggregates, we found $ \tau_\infty = (0.030 \pm 0.007) \, \rm Pa$ at $ 2 \,\mathrm{mbar}$, that is, at a pressure where $f_\mathrm{c} = 1$ for the dust aggregates. This data point is not integrated into Fig. \ref{results} because it is only relevant here for determining $\alpha$. Since $\alpha$ is the friction coefficient, it is sample specific, but does not depend on gravity. By using Eq. \ref{shortcut}, we obtain $\alpha = 0.035 \pm 0.007$. When this is known, Eq. \ref{eq:tau_erosion_final} can be fit for $\gamma_\mathrm{e}$ and $\beta$.
In summary, we find
\begin{eqnarray}
\label{eq:werte}
\nonumber \alpha = (3.5 \, \pm \, 0.7) \cdot 10^{-2}\\
 \gamma_\mathrm{e}= (7.8 \, \pm \, 1.3) \cdot 10^{-5} \,\mathrm{J}\,\mathrm{m}^{-2}\\ 
\nonumber \beta = 3.48 \, \pm \, 0.55. 
\end{eqnarray}
The data for the dust aggregates can clearly be well described by the equation found for glass beads. The validity of this equation is therefore extended to larger grains and lower pressures.
The specific material constants are slightly different. \citet{Demirci2020} reported higher values for $\beta$ with increasing particle size, and our data seem to follow this trend as well. We recall, however, that we used dust aggregates in a dust-aggregate bed now, which might have slightly different aerodynamic properties \citep{Schneider2021}. 

It might have been expected that dust would require higher shear stress to become entrained. That this is not the case depends on the somewhat lower density, but the effective surface energy also shows that the sticking is not higher. The history of their generation, terminated at the bouncing barrier, makes dust aggregate pebbles as 
little sticky as solid monolithic pebbles would be.

It might further be noted that these dust pebbles and their assemblies are also rather robust. Dust aggregates entrained in the gas flow and impacting on the granular bed sometimes later just produce splashes that are very similar to splashes of glass beads or other solids. While some fragmentation can take place at high-velocity impacts, it does not seem to change the morphology of the granular bed. At least erosion can proceed normally afterward. This is important to note as in protoplanetary disks, eroded pebbles can become projectiles for other planetesimals. Our experiments suggest that this might not influence gas-drag-based erosion, but rather adds to erosion, although this would be a different topic.

\section{Forbidden orbits for planetesimals}
\label{sec:astrophysicalapplication}

As mentioned above, recent works concluded that pebble-pile planetesimals might not persist on every orbit. Depending on the eccentricity and disk model, the inner regions are excluded as these bodies are readily eroded. With our new values for the erosion of dust aggregate pebbles, we update these erosion zones here.

To do this, we have to compare the threshold shear stress $ \tau_\mathrm{t} $ above which a planetesimal is eroded with a maximum wall shear stress $ \tau_\mathrm{w} $ that acts on it at a certain distance from the star.
Equation \ref{eq:tau_erosion_final}, together with the experimentally determined material-specific values (Eq. \ref{eq:werte}), provides us with a way to calculate $ \tau_\mathrm{t} $. It can be translated into a dependence on the semi-major axis of the planetesimal orbit $ R_\mathrm{o} $. Even though $ R_\mathrm{o} $ does not appear directly in Eq. \ref{eq:tau_erosion_final}, it is implicitly included via the spatial dependence of the gas density $ \rho(R_\mathrm{o})$ (see Eq. \ref{eq:rho_gas}). The planetesimal radius $ r $ and the pebble size $ d $ can be set as parameters.\\

A few equations are currently discussed in the literature for
the maximum wall shear stress  $ \tau_\mathrm{w} $ that a body with radius $ r $ experiences in a gas flow. They might all be applicable here.  For an orbit with semi-major axis $ R_\mathrm{o} $ and eccentricity $ e, $ we therefore consider three different estimates in the following. For the first two, we can simply use Eq. \ref{eq:shearing2}, but with the relative velocity $ v $ between planetesimal and the surrounding gas and the thickness of the laminar boundary layer around the flowed body $ h_\mathrm{B} $ 
\begin{equation}
    \tau_\mathrm{w} = \eta_\mathrm{h} \frac{v}{h_\mathrm{B}}.
    \label{eq:tau_w general}
\end{equation}
At this point, we use the values for hydrogen instead of air for all gas-specific constants, that is, $ R_\mathrm{g,h} = 4124.2 \,\mathrm{J} \,\mathrm{kg^{-1}}\,\mathrm{K^{-1}} $ and $ d_\mathrm{m,h} = 2.76 \cdot 10^{-10} \,\mathrm{m} $. The dynamic viscosity of hydrogen $ \eta_\mathrm{h}$ depends on the square root of the temperature   
\begin{equation}
     \eta_\mathrm{h} = \eta_\mathrm{0}  \sqrt{\frac{T}{280 \, \rm K}}
    \label{eq:eta_h}
,\end{equation}
\noindent with $ \eta_\mathrm{0} =  8.9 \, \rm \mu Pa \, s $ being the dynamic viscosity at $ T = 280 \, \rm K$ \citep{May2007}. The temperature in the protoplanetary disk is 
\begin{equation}
     T = T_\mathrm{0}   \sqrt{\frac{R_\mathrm{o}}{1 \, \rm au}}
    \label{eq:Temperature}
,\end{equation}
\noindent with $ T_\mathrm{0} = 280 \, \rm K $ \citep{Hayashi1981}. As relative gas velocity $ v, $ we assumed
\begin{equation}
    v = v_\mathrm{g} + \Delta v = v_\mathrm{g} + \left( \sqrt{\frac{GM}{R_\mathrm{o}}\left(\frac{1+e}{1-e}\right)}-\sqrt{\frac{GM}{R_\mathrm{o}}} \right)
    \label{eq:velocity}
,\end{equation}
where we took $ v_\mathrm{g} = 54 \, \rm m/s $ as the minimum gas velocity on a circular orbit around the star due to the pressure gradient \citep{Weidenschilling1977}. The constants $ G $ and $ M $ are the gravitational constant and solar mass, respectively. If the orbit is not circular but eccentric, the headwind can increase significantly. We took the periastron as the position of erosion because the velocity is highest at this point. $ \Delta v $ in Eq. \ref{eq:velocity} takes this into account by describing the difference between the orbital velocity at periastron for a given eccentricity and for zero eccentricity.\\
The thickness of the laminar boundary layer around a planetesimal is not well defined, and different approaches were used to estimate it \citep{Demirci2020c,Cedenblad2021,Schlichting2006}. We used the following two boundary-layer-approximations for the calculation of $ \tau_\mathrm{w} $:
\begin{equation}
   \begin{cases}
    h^{\mathrm{Ced}}_\mathrm{B} = \sqrt{\frac{2 r \eta_\mathrm{h} }{\rho_\mathrm{g} v}}, \ \ \text{Cedenblad et al. (2021)} \\[2\jot]
    h^{\mathrm{Sch}}_\mathrm{B} = 5\sqrt{\frac{2 r \eta_\mathrm{h} }{\rho_\mathrm{g} v}}, \ \ \text{Schlichting} \ \& \ \text{Gersten (2006).}
  \end{cases}
  \label{eq:boundary layer}
\end{equation}
In a third approach, the wall shear stress is calculated by an expression using the threshold friction velocity $ u^* $, similar to what was done by \cite{Demirci2020}:
\begin{equation}
    \tau_\mathrm{w} = \rho_\mathrm{g} u^{*2} = \rho_\mathrm{g} \biggl[ 0.702  v  \left( \text{Log}_{10} \frac{2r \rho_\mathrm{g} v}{\eta_\mathrm{h}}\right)^{-1.59} \biggr]^2.
    \label{eq:greeley}
\end{equation}
The empirical equation for $ u^* $ was developed by \cite{Greeley1980} on the basis of wind-tunnel experiments. For the final calculation of $ \tau_\mathrm{w} $ as well as $ \tau_\mathrm{t}$, an expression for the gas density $ \rho_\mathrm{g} $ and the gravitational acceleration $ g $ on the planetesimal is still needed. For gravity, we took
\begin{equation}
    g=\frac{4}{3} \pi G \rho_\mathrm{pla} r
    \label{eq:gravity}
,\end{equation}
with $\rho_\mathrm{pla} = 538 \, \rm kg/m^3 $ \citep{Paetzold2018}. As gas density, we took a power law according to the minimum-mass solar nebula \citep{Hayashi1981},
\begin{equation}
    \rho_\mathrm{g} = \rho_\mathrm{0} \left( \frac{R_\mathrm{o}}{1 \, \mathrm{au}}\right)^{-\frac{11}{4}}
    \label{eq:rho_gas}
,\end{equation}
with $ \rho_\mathrm{0} = 1.4 \cdot 10^{-6} \, \rm kg/m^3$. When we take all this together, we have the tools for calculating $ \tau_\mathrm{w} $ and $ \tau_\mathrm{t}$ as a function of the orbit parameter $ R_\mathrm{o} $ with the planetesimal radius $ r $, the pebble size $ d $, and the eccentricity of the orbit $ e $ as selectable parameters. With this, we can set up the following requirement for the stability of a planetesimal: 
\begin{equation}
    \frac{\tau_\mathrm{t}(R_\mathrm{o},r,d)}{\tau_\mathrm{w}(R_\mathrm{o},r,e)} \geq 1
    \label{eq:tau_requirement}
,\end{equation}
which allows us to make predictions on stable and unstable orbits.
\begin{figure}
        \centering
        \includegraphics[width=\columnwidth]{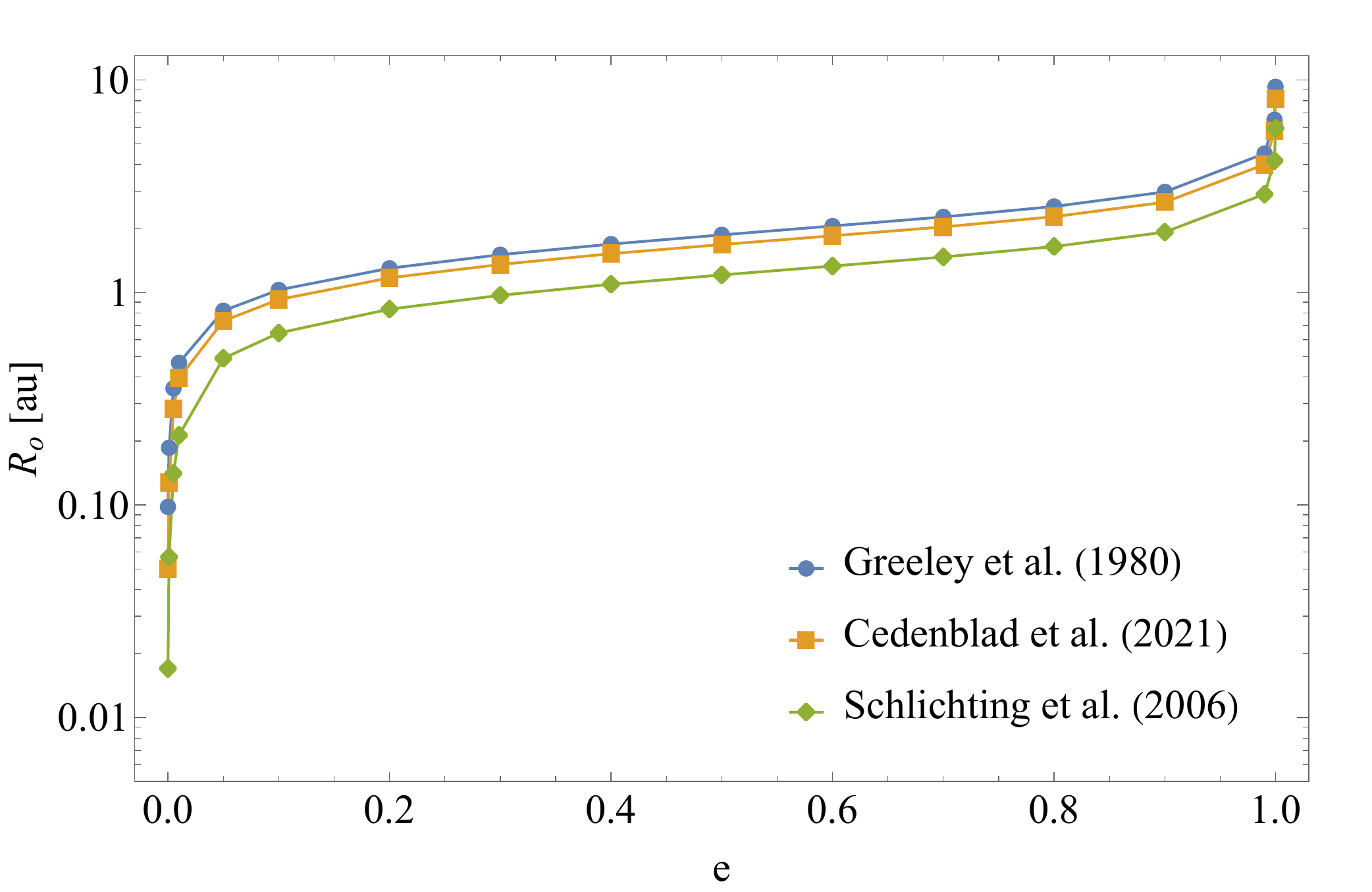}
        \caption{\label{fig:R_orbit(e)} Minimum orbits (semi-major axis $ R_\mathrm{o} $) for given eccentricities $ e $ at which a 2 km sized planetesimal composed of 1 mm pebbles is stable, i.e., $ \tau_\mathrm{t} / \tau_\mathrm{w} = 1 $ for three different options for calculating $\tau_\mathrm{w}$. At smaller orbits, the planetesimal is dismantled.}
\end{figure}
For a 2 km body with 1 mm pebbles, Fig. \ref{fig:R_orbit(e)} shows the minimum safe orbits for different eccentricities $ e $, that is, the value of the semi-major axis $ R_\mathrm{o} $ for which the ratio of $ \tau_\mathrm{t} $ and $ \tau_\mathrm{w}$ equals one. For lower values of $ R_\mathrm{o} $, the shear stress on the planetesimal is greater than the threshold shear stress. It therefore becomes dismantled by the headwind. Figure \ref{fig:R_orbit(e)} therefore illustrates a forbidden zone for planetesimals that extends up to 1 au even at low eccentricities of $ 0.1 $. If the eccentricities are higher, the threshold between forbidden and safe zone is displaced outward up to 3 au. In a forbidden zone, every pebble pile smaller than the respective planetesimal size is also instable down to a size at which particles decouple from the gas. Furthermore, these zones are somewhat larger than in earlier work now because the dust aggregate pebbles are not bound as strongly.

\section{Planetesimals in later evolutionary phases}

\begin{figure}
        \centering
        \includegraphics[width=\columnwidth]{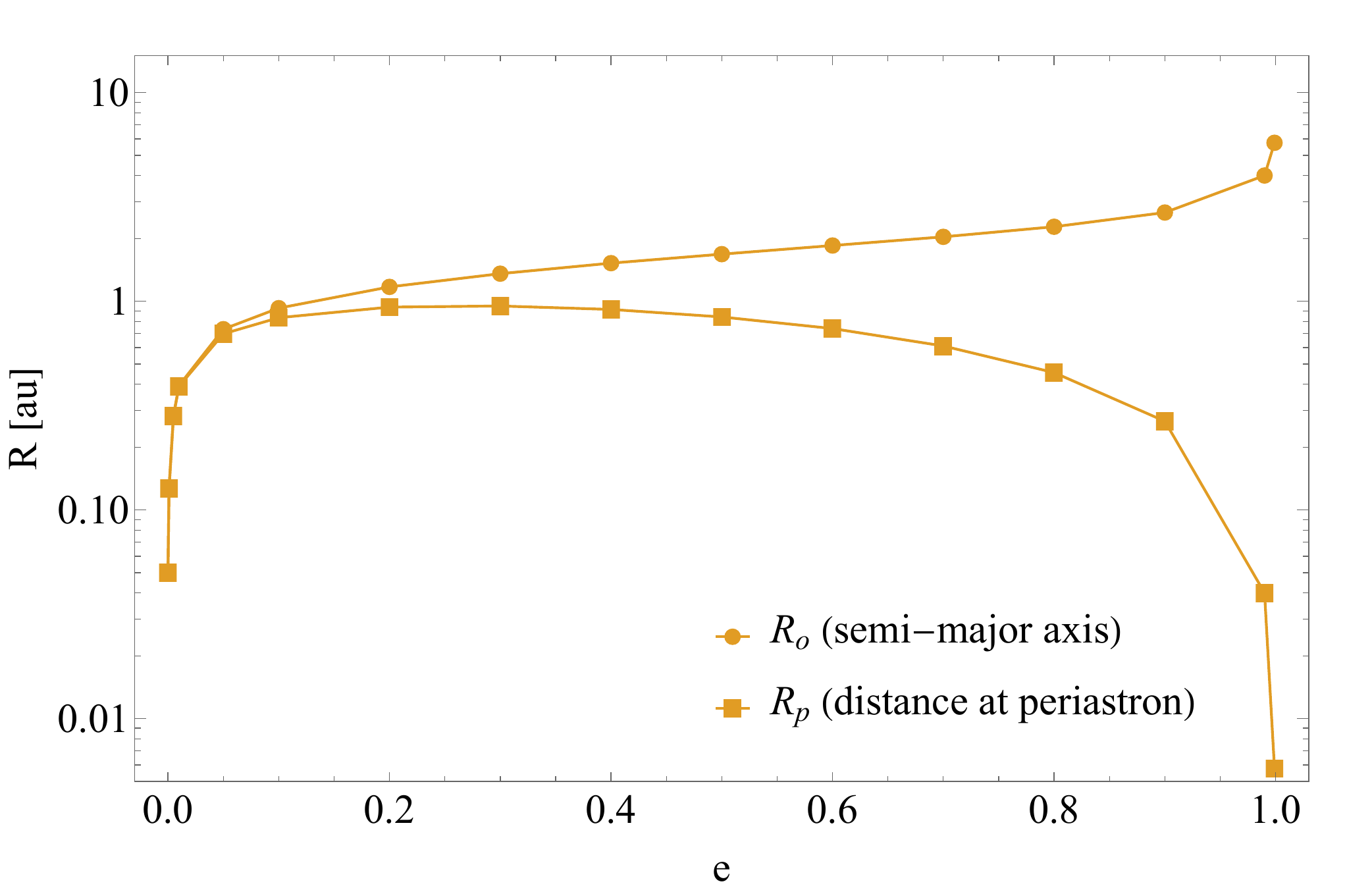}
        \caption{\label{R_periastron_R_orbit} Comparison of minimum stable orbits with the corresponding distances from the star at which most material is eroded. Upper line (circles): semi-major axis $ R_\mathrm{o} $ of an eccentric planetesimal prone to erosion. Lower line (squares): periastron distance $ R_\mathrm{p} $ at which pebbles are released (2 km planetesimal and 1 mm pebbles; boundary layer from \citet{Cedenblad2021}).}
\end{figure}

In early phases of planet formation, the first forming planetesimals are likely on circular orbits and are only instable in a region close to the star. In later phases of planet formation, planetesimals are still present, but now planets can stir them up into eccentric orbits. These planets are thought to grow by accreting pebbles or planetesimals \citep{Bitsch2019, Brugger2020}. The fate of eccentric planetesimals is therefore relevant. If a planetesimal that encounters a planet is neither destroyed nor accreted, but scattered into an eccentric orbit, it might be dissolved into pebbles farther inward. 
To get an idea of the absolute scales, Fig. \ref{R_periastron_R_orbit} shows an example of the periastron distance where pebbles are dumped by a planetesimal on an instable orbit.

For a planet outside of 1 au, these scattered planetesimals are lost as a source to accrete from. For a planet inside of 1 au, the generated pebbles are a new source of material.

\begin{figure}
        \centering
        \includegraphics[width=\columnwidth]{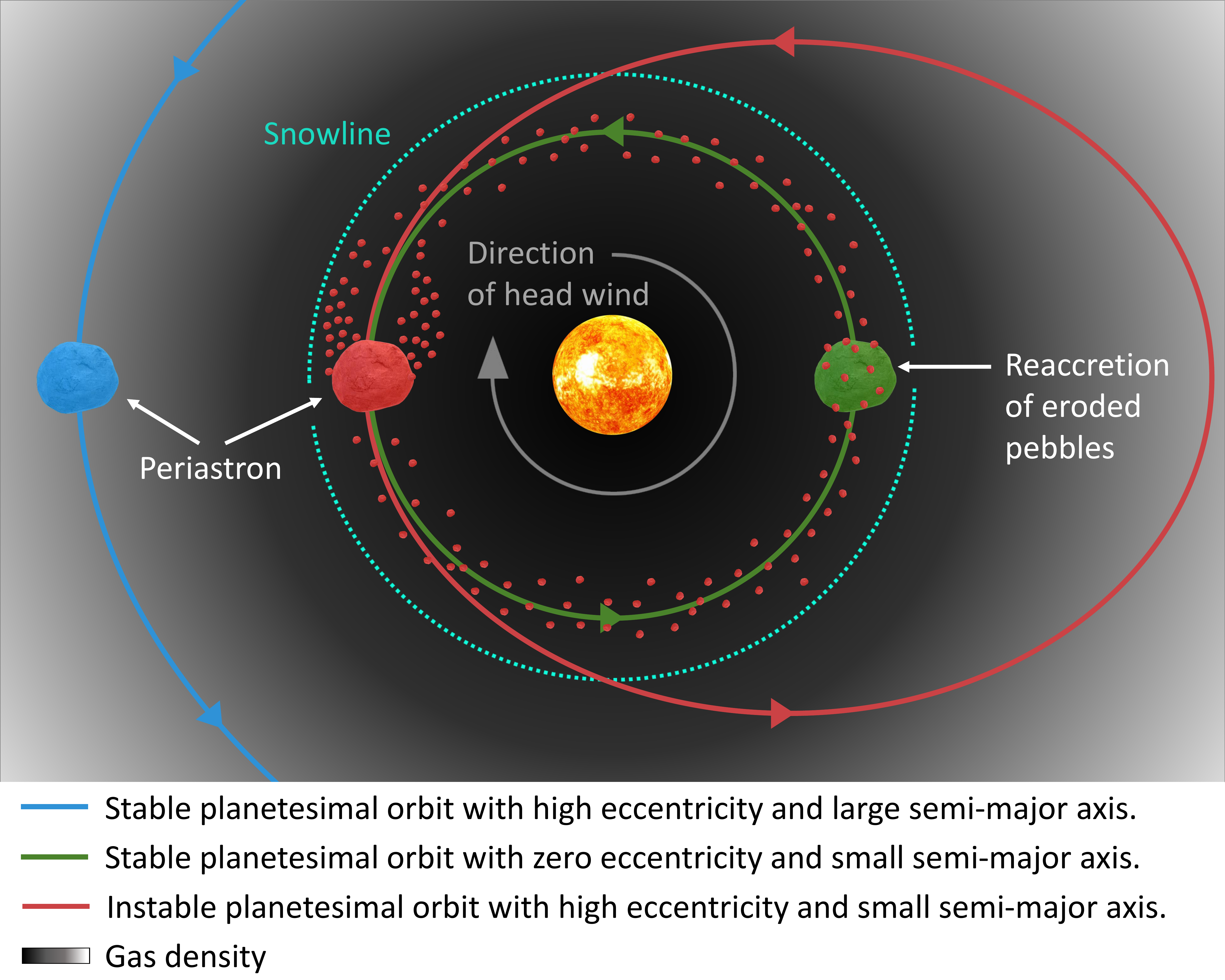}
        \caption{\label{fig: schematic} Schematic illustration of different planetesimal orbits. Planetesimals are stable in the outer disk (blue) and on circular orbits within 1 au (green). Eccentric planetesimals (red) will be lost for any accreting planet outside of 1 au, but provide pebbles for accretion onto a planet inside 1 au. If the snowline is around 1 au, water might also be efficiently transferred inward.}
\end{figure}

If the snowline were in between, and if icy planetesimals could be disassembled in the same way, then water could also be transported more readily from beyond the snowline to a place farther inside of the snowline. Figure \ref{fig: schematic} shows a schematic of these ideas. 

\section{Conclusions}
\label{sec:conclusion}

In laboratory experiments on parabolic flights, we extended earlier findings on the erosion of pebble-pile bodies in protoplanetary disks. We used $\sim 1 \, \rm mm$ dust aggregates as pebbles now and were able to decrease the ambient pressure for which we find erosion by another order of magnitude to 1 Pa. While the gravity level is still higher than on planetesimals, the extrapolation to protoplanetary disks becomes more reliable. In general, dust aggregate pebbles were found to be more susceptible to erosion than the glass beads used before, that is, they are less dense and as little sticky as glass beads.

In the scenario in which planetesimals form through a gentle gravitational collapse of a pebble cloud, some zones are ruled out for their existence. High-eccentricity planetesimals can only survive erosion by the head wind of a protoplanetary disk outside of about 1 to 3 au, but even planetesimals on circular orbits are not stable inside of 0.01 to 0.1 au.

Therefore, in one way or another, erosion sorts out different populations of planetesimals at different locations and provides matter for generating the next generation planetesimals or for accretion onto planets farther inward. This might also be a way of transporting water inward. The architecture of planetary systems will be shaped by the destruction of pebble-pile planetesimals and their reassembly. To what end this shaping is used remains to be seen, but as erosion can decide whether a planetesimal with certain orbital parameters exists or not, it should make a significant difference. 

\begin{acknowledgements}
      This project is funded by DLR space administration with funds provided by the BMWK under grant 50 WM 2140. K.J., F.C.O., and T.B. are funded by grants 50 WM 1943, 50 WM 2142, and 50 WM 2049, respectively. M.F. is funded by DFG TE 890/7-1.
\end{acknowledgements}

%-------------------------------------------------------------------

\bibliographystyle{aa} % style aa.bst
\bibliography{forbiddenplanetesimals} % your references Yourfile.bib

\end{document}